# Knowledge Discovery from Social Media using Big Data provided Sentiment Analysis (SoMABiT)

**Mahdi Bohlouli**
Institute of Knowledge Based Systems, Department of Electrical Engineering and Computer Science, University of Siegen, Germany

**Jens Dalter**
Department of Business Information Systems, University of Siegen, Germany

**Mareike Dornhöfer**
Institute of Knowledge Based Systems, Department of Electrical Engineering and Computer Science, University of Siegen, Germany

**Johannes Zenkert**
Institute of Knowledge Based Systems, Department of Electrical Engineering and Computer Science, University of Siegen, Germany

**Madjid Fathi**
Institute of Knowledge Based Systems, Department of Electrical Engineering and Computer Science, University of Siegen, Germany

**Abstract**
In today's competitive business world, being aware of the customer needs and market oriented production is a key success factor for industries. To this aim, the use of efficient analytic algorithms ensures a better understanding of the customer feedback and improves the next generation of products. Accordingly, the dramatic increase of using social media in the daily life provides beneficial sources for market analytics. But how traditional analytic algorithms and methods can scale up for such disparate and multi-structured data sources is the main challenge in this regard. This paper presents and discusses the technological and scientific focus of the SoMABiT as a social media analysis platform using big data technology. Sentiment analysis has been employed in order to discover knowledge from social media. The use of MapReduce and developing a distributed algorithm towards an integrated platform which can scale for any data volume and provide a social media driven knowledge is the main novelty of the proposed concept in comparison to the state-of-the-art technologies.

## 1. Introduction

Scalable Decision Making (SDM) in data-intensive applications is a key challenge for enterprises dealing with big data. Complexity and expansion of resources and streams as well as velocity and variety of the data have led to difficulties in enterprise asset management [1]. In this context, scalable algorithms and semantic technologies for dynamic, economic and efficient management of information could facilitate the transition from current challenging situations into a scalable and adaptive decision support systems era in enterprise practices. As an example, in the air traffic management with data intensive activities such as detection of a conflict between two arriving aircraft, a key challenge is to semantically integrate, analyze and interpret data such as trajectory with flight plans, etc., so that they can be used directly for negotiation and resolution of conflicts [2]. The most important factors in such a scenario are time, cost and scalability.

**Corresponding author:**
Mahdi Bohlouli, Institute of Knowledge Based Systems, Department of Electrical Engineering and Computer Science, University of Siegen. Hoelderlinstr. 3, D-57076, Siegen, Germany. Email: mahdi.bohlouli@uni-siegen.de

Applications should not only be nearly on-time and cost effective, but also adequate to scale up by enhancing data and processing demands and access via different types of platforms anytime.

On the other hand, social media provides a massive amount of data, especially user generated-content that can be used for opinion mining for a wide variety of purposes. According to Nielsen's report, the consumer-generated product reviews and ratings are the most preferred source of information between social media users [3]. Accordingly, social media analysis provides product improvement recommendations as well as smart and novel ideas for the next generation and new products. This also provides more efficient marketing methods for enterprises in today's competitive business world. It is not limited only to industry and marketing oriented aspects, but also to social, medical and political goals.

However, the main challenge here is to find out how to collect the data from social media, which data sources can be useful for specific goals, how to analyze collected data and discover useful knowledge from sources and how to provide scalable algorithms in order to proceed with the large volume and variety of data sources provided in social media. For instance, according to the reports released in February 2015, every day, Twitter produces about 500 million tweets[1], Facebook produces[2] 2.5 billion pieces of content (100 terabytes) and 144.000 hours of videos are being uploaded to the Youtube[3].

Managers and key decision makers in various industrial and/or business sectors involve continuous supervision of huge amounts of information, which have to be collected from social media and analyzed for predictions, judgments, evaluations, strategic plans and actions. In most cases, the decisions that have to be made are subject to strict restrictions regarding available resources and requested response times. Moreover, decision makers usually encounter sudden, unexpected and urgent events that can easily lead to dangerous situations that may threaten the safety and the reliability of the whole system. The issue of how to scale up or down for acceleration or deceleration imposes an additional complexity in today's traditional decision making methods [4]. This issue becomes more challenging when there are not any particular estimates of future growth for data volumes and service demands.

Managing huge amounts of heterogeneous data has recently emerged as a key challenge in many computing applications. In addition to the traditional Relational Database Management Systems (RDBMS), so-called NoSQL databases have appeared as high performance alternatives, providing document-oriented storage for semi-structured or unstructured data [5]. These databases can also be deployed to many nodes and allow adjustable redundancy levels as required by the application. However, the right choice of database management system and its correct parameterization according to the data as well as the data processing requirements of a specific application are not yet fully understood in a big data era [4]. Many aspects have to be taken into account, like the type of data assets, the access patterns in the data, the desired level of redundancy/availability, the isolation level of distributed nodes, and many more.

The integration of heterogeneous data sources presents another key challenge. Current approaches of joining diverse data sources and creating an abstraction layer for unified data access are often the result of an ad-hoc approach [6]. A comprehensive methodology for creating data federations over diverse data sources, which is applicable for different domains, is still missing. This is particularly critical when static and dynamic data sources have to be combined for creating new insights from the data specially through social media sources.

NoSQL database management systems are characterized by not using SQL as a query language - or, at least, not using fully functional structured queries. Mostly they do not offer relational operators like JOIN and generally do not provide full ACID (Atomicity, Consistency, Isolation and Durability) guarantees [7] in terms of atomic transactions, consistent database states, transactional isolation, and durability of persistent data. On the other hand, NoSQL databases offer good performance and horizontal scaling across the nodes of a cluster. As such, they are well suited for web-scale applications and other big data domains, where the efficient storage and access of huge data volumes is more important than transactional consistency.

It is imperative to clearly define the strengths and weaknesses of NoSQL technology, where databases are not relational and have no fixed data scheme, complex relations or joins. The common denominator of the majority of NoSQL databases is that they are optimized for large or massive data-store scaling, i.e. they are supposed to scale more efficiently and smoothly than RDBMSs by spreading the processing and storage load over a multitude of affordable server systems. On the other hand, relational database management systems - SQL databases - scale up by using ever faster and memory/disk rich high-end server hardware [4]. HBase technology [8], as a NoSQL database solution, which is a part of the Hadoop ecosystem, has been used in the data layer. One of the major advantages in using HBase is its distributed and scalable database support with random and real-time reads/writes.

The article is structured in five sections. The first section provides an introduction and motivation of the research. The next section (Sec. 2) reviews state-of-the-art technologies, tools and findings from different perspectives that relate to this research. The section focuses on big data, social media analysis and semantic technology areas and summarizes how the integrated use of these technologies could discover knowledge from social media and improve the quality of products. Section 3 discusses about the SoMABiT concept and provides requirement analysis as well as the list of the data sources that have been used in the system. The realization, scientific background, evaluation and verification of results are afterwards discussed in the section 4. The conclusion and future works are addressed in section 5 which summarized the findings and provides an outlook on the future work of the project.

## 2. State-of-the-Art technologies

In this section, a brief overview of key tools and technologies are given by categorizing them into singular subsections. The authors would like to first study big data technologies and continue by stating the advancements in semantic technologies and social media marketing and analysis, which will provide a required background for the target multidisciplinary research.

### *2.1. Big data analytics and visualization*

From a conceptual perspective, NoSQL database can be roughly divided into four different categories. These are the wide column stores/column families, document stores, key/value/tuple stores and graph databases [4]. The key/value/tuple stores operate on the principle of storing records in a form of keys and values. The values can represent any kind of data and type [4]. This has the advantage of a fast data access and a simple data management. It is also possible to keep very large data sets ready and distribute them to different nodes due to better scalability. A disadvantage exists in the use of more complex queries, because the same options as for relational systems aren't given. The wide column stores/column families are built on the principle of key/value/tuple stores, complemented by the possibility of expanding individual columns to other key-value pairs [14]. The document stores are used to store structured data, such as JSON (JavaScript Object Notation) files. They are used when only certain predefined file types must be stored in large quantity. The graph databases represent another way in which data can be stored. These are treelike structured data records which are interconnected. The data in the form of geo-information is an example of this type. This results mainly from the ability to manage large amounts of data, which may be unstructured. For this reason the use of relational databases is not endorsed, on the basis of the fact that only pre-defined data types can be stored.

After publication of the MapReduce algorithm and Google File System (GFS) [10] [11], the Hadoop implementation [12] has been initiated by Yahoo and continued as an open source Apache project. Its main components are the MapReduce implementation and the Hadoop Distributed File System (HDFS). One of its key features is fault tolerance and the ability to run on clusters of unreliable commodity hardware.

Apache Hadoop provides a programming model that is suitable for the parallel processing of large data sets. Essentially, it consists of the two phases, that are called "map" and "reduce" which are applied particularly to text data. In the "map" phase, the whole data gets iterated and processed into a specific format according to the desired result. The data records themselves are presented as key-value pairs. In the following "reduce" phase, all key-value pairs are merged. This means that values with the same keys are combined. Thus, the total amount is reduced to the data sets and by key values [12].

HBase is an open source and integrated database with Hadoop/HDFS [13]. It is distributed, column-oriented, distributed configurable, scalable and easy to access through MapReduce [4]. The main focus of HBase is to support big tables with hundreds of billions of rows and columns based on Google BigTable. HBase provides higher performance when it is being accessed through many distributed clients. It has a slow latency of individual transactions due to the network traffics in distributed environments. HBase supports structured data sources, but porting the applications have been built using RDBMS to HBase is not possible. The key in the HBase table represents a string that points to the row with the corresponding data. In turn a column contains also keys and values. The values are stored as byte arrays and thus have no fixed data type. Individual columns can be grouped into column families.

Mahout is an open source software project hosted by the Apache foundation [14]. It provides a machine learning library on top of Hadoop, with the goal to provide machine learning algorithms that are scalable for large amounts of data. The development has been initiated with Chu et al. [15] and several dozens of algorithms have been implemented up to now for data clustering, data classification, pattern mining, dimension reduction and some others. All algorithms are written in Java and make use of the Hadoop platform.

Cloudera is an Apache-Hadoop based software suite and is presented in different bundles [16]. The target platform in this work is Cloudera Enterprise Free. The software supports CDH (Cloudera's Distribution, including Apache Hadoop), HDFS, MapReduce and Hadoop developments in the cluster as well as user and workflow management by incorporating different services. CDH is the open-source Hadoop distribution of Cloudera. The core component of Cloudera, which supports the management of CDH by sets of standards and delivering granular visibility over every part of CDH [16], is Cloudera Manager. It offers particularly far-reaching administrative capabilities of a cluster system. It also includes features for the diagnosis and monitoring of cluster systems.

Lorica [17] issued the software development process as a key element for efficient monitoring and visualization of social media data. The main concern here is large volumes of these datasets. He reported 500 million tweets daily at the date of publication which is even more nowadays. In this regard, flexible and interactive software development frameworks facilitate the visualization of social media data. Big data visualization is being used in a wide variety of sectors and applications such as medicine, military services as well as national security, public service and financial areas. For instance, DENVIS as an end-to-end solution, which is proposed in [18], uses MapReduce to manage, mine, visualize, and analyze large dental imaging data in two main parts: "data driven image analysis modules triggered by imaging data acquisition that exploit parallel MapReduce tasks and ingest visualization archive into a distributed NoSQL store, and user driven modules that allow investigative analysis at run time". Furthermore, ConnectomeExplorer [19] applied big data visualization in the field of neuroscience which visualizes large volumemetric electron microscopy (EM) datasets in connectomics research. The dataset used in this research is about 1 terabyte EM data as well as 750 GB segmentation data consisting 4,000 segmenting structures and 1,000 synapses. In addition, Ng and Qu [20] and Cui et al. [21] used big data technology in visualization of financial data and news streams, respectively. Such applied and practical researches show the application and importance of big data visualization in a wide variety of applications. There are even more researches and funded projects in the areas of geosciences, weather forcasting as well as astonomy. Traditional methods and algorithms provide similar facilities as well, but they lack efficient and scalable algorithms for visualization and analysis of such large and unstructured data due to the 3Vs concept (Value, Volume, Velocity).

## *2.2. Semantic technologies*

Semantic technologies are an important factor in the analysis components of today's data intensive applications and network structures. Dengel [22] defines semantic technologies as a tool for analyzing linguistically phrases in a descriptive and formalized way, thus improving the understanding between communication partners and the execution of actions originated by computers. The main approach for the connection of internet data is the semantic web[4], bridging semantic features in different layers, which offer many standardized technologies and protocols for representing and deriving data. Extensible Markup Language (XML), Resource Description Framework (RDF) or Web Ontology Language (OWL), explained among others by Breitman et al. [23], build the most important and standardized semantic layers. These technologies have different levels of representing semantics.

The XML is a "general-purpose markup language, designed to describe structured documents" [23]. The representation of semantic connections is realized in tree like structures, while the RDF allows the creation of interconnections between different resources on the web in a graph structure, thus allowing the representation of connections between objects not necessarily arranged in a hierarchy [24]. Fensel [24] defines RDF as "an infrastructure that enables the encoding, exchange, and use of structured metadata". RDF representations of websites (e.g. DBPedia, the semantic representation of Wikipedia) may be queried via SPARQL. The SPARQL Protocol and RDF Query Language apply a similar syntax then SQL, the database querying language, thus allowing the extraction of knowledge of RDF structures [22]. RDF provides a means of creating basic ontologies, OWL extends these possibilities to building more complex ontologies [23].

The core characteristics of an ontology are that it is formalized with well-defined semantics and contains a shared understanding within a community [24]. An ontology may therefore be defined as "shared models of some domain that encode a view which is common to a set of different parties" [25]. In the semantic web architecture[4] of the W3[5], ontologies are queried via rules [26]. Lee et al. [27] used probability for representing a model in order to retrieve documents with a high degree of the semantic relevance. The use of the semantic technology in this research provides enriched semantics of the user needs in the semantic search and recommendation.

Web Rules may have different requirements towards classical rule based systems which are applied in the context of expert systems. These requirements originate from the different structures of the web content, where often lots of the provided facts are likely to be changed and in consequence may draw a lot of irrelevant conclusions [26]. Today there are different web rule languages or rule engines available. The semantic social networks model that introduced by Jung and Euzenat in [28] consists of three layers as of social network (network of the people), ontology network and a network of concept in ontologies. They used relationship analysis in the social networks by means of the similarity measures for emergence of consensus ontologies. Jung [29] also proposed a concept of integrating the social network for context fusion in mobile service platforms. In this context, he used social networks integration for providing a context aware platform for service providers.

Next to these formalized standards of the semantic web stack[4], applied semantic technologies in the web context are for example Linked Open Data (LOD[6], based on RDF triples) for connecting data sources on the web, semantically enhanced search engines for improvement of search results or semantic web services, where "normal" web services are semantically enhanced and allow for automated execution processes [22].

*2.3. Social media marketing and analysis*

Social media is a phenomenon which emerged parallel or in the wake of the Web 2.0 development and grounds on Web 2.0 technologies [30]. The main applications of social media are social networks (for private and business sector), blogs, micro-blogs, wikis, content sharing, bookmarking, rating and ranking platforms as well as crowdsourcing. These applications possess different purposes and characteristics of social communication and interaction [31, 32].

Especially companies address the provided services as tools for many different areas like social media marketing, market research, public relations, sales, services, support as well as product development, recruiting, and internal communications [31]. Furthermore, they apply it for promoting or visualizing the image of their company [33]. The customers on the other hand use it for comparing, rating, ranking, or giving feedback about products. This again influences their decision about buying a specific article or not [34].

To establish a close relation to already or potential customers, companies freely use social media tools for marketing campaigns of their products and/or interacting with the customers. Social media marketing increases the awareness of the company, helps in establishing a positive image and a closer connection to the customers, while at the same time promoting new products or brands [33]. Hettler [34] also points this fact as a new form of interaction, which is a source for market research and consequential decision making. Other companies go even one step further and incorporate customers in their innovation or development processes, e.g. in the form of crowdsourcing [34].

Hwang et al. [35] discussed how information is being created, shared and "diffused by interactions among users through online social media". However, they proposed a diffusion pattern process which can be recognized in two different types of analytic methods with macroscopic and microscopic approaches. According to [35], the diffusion patterns in the social media platform, Twitter, are being influenced by events, emotional words, and the number of followers of each user.

Hassler [36] defines social media analytics as a way to monitor or analyze all third-user generated data in the web about the own company, products or brands. In this context, social media data is "vast, noisy, distributed, unstructured, and dynamic" [32] which demands special requirements for structuring and analyzing. Industry and science develop and improve social media analysis and monitoring services to support companies in monitoring their social media activities. Additionally, the sentiment or opinion of the customers about the products or company's services is a deciding factor for further product improvement, as it offers a source of direct feedback from the customer, without a formal questionnaire or interview. If a customer is not satisfied with the product, he may easily post a message via social network, where he expresses his opinion which allows tools for sentiment analysis to gather the positive or negative feedback. Due to the fact that other people react to such posts by, for instance "like" or "retweet" over different social media channels, the company may receive indirect positive or negative marketing for the products, as customers "are talking about them".

In literature, there are different terminologies for analyzing social media like social analytics (e.g. [37]), social media analytics (e.g. [36]), social media monitoring (e.g. [34]) or social network analysis, while the last one focuses more on the connections between the users rather than on the content itself. Lovett [37] defines four core metrics for measuring social media: "foundational measures", which "include interaction, engagement, influence, advocates and impact" of social media actions, "business value metrics", "outcome metrics (KPIs)", which have to be defined independently by each company and finally "counting metrics", which include the counting of followers, page views or comments to a social media page of the company. Mozafari et al. proposed a social behavioral information diffusion model [38] based on social network analysis. Their model inspired from information propagation between neighbors and considers the effect of mainstream media like TV and radio. In [39], the authors used social network in order to automate village paradigm by selecting the candidates through social networks. Vilares et al. [40] analyzed social media (Twitter messages) from a linguistical perspective in order to mine the most relevant messages about the specific issue.

The perception of opinions in social media is particularly important because reviews from other consumers or friends have a special relevance. In addition, many customers run an internet research about the products they wish to purchase. These reviews from other consumers are considered to be particularly significant. It begs the question of how to evaluate such large amounts of data, arising through social media. The use of traditional business intelligence systems is mostly not useful because they can be overwhelmed technically and conceptually with the volume and variety of data sources. This causes because of the exponential growth, complexity and ever-changing needs of the data. Through the use of big data systems, competitive advantages can arise for businesses, especially if the results offer the possibility of adapting the business models. Through the use of already available data, no data collection in the form of questionnaires or interviews must be conducted. This information can be obtained free of charge (within the meaning of the survey). Analyzing the data in a short time supports companies to react quickly based on the customer needs and market changes.

*2.4. Interconnected big data based semantic social media analysis*

In regards of the cloud deployed integrated semantic social media analysis, there are different approaches as of semantically enhanced cloud structures (e.g. [41]), semantics used for managing the cloud (e.g. [36]), semantic representations or enrichment of big data. Haase et al. [42] describe an eCloudManager architecture, where the core component is a semantic data store for connecting the content from different data sources of the data integration layer. Sheth and Ranabahu [41] argue applicability, especially to supplement interoperability in the cloud space for semantic models. The data from social media is multi-structured (Variety), very large scale and distributed (Volume) and sometimes needs to be analyzed in a real-time manner (Velocity). For such data sources with these specifications, this is known as the 3Vs aspect in the big data domain [4]. The authors believe that the fourth V, which is "Value", is needed for making sense of social media analysis. The process of knowledge extraction from big data analysis conducts value creation. In this regard, semantic technologies play a key role in understanding the user queries and semantic analysis of requirements. The whole platform should be deployed as a cloud-based solution due to the limitations with in-house computing resources at peak times as well as some more advantages that have been stated earlier.

*2.5. Sentiment analysis*

The sentiment analysis is an increasingly popular text mining method to determine the opinion of a text. It is also often referred to as opinion mining. The sentiment analysis uses individual elements of a text on different text levels, like whole document, paragraph, sentence or only a text window, and tries to correctly determine and assign the particular mood and emotions to the respective entities, also called objects [43]. The challenge of this analysis is to replicate a human process - the understanding of text information, assignment of the individual polarized information and interpretation of the human mind. Basically, the problem arises, that computers do not own interpretive skills, and can therefore only learn to understand complex relationships based on models with their artificial intelligence. Because of this limitation it is necessary for the sentiment analysis to process the unstructured data into a form which is structured and machine-readable. In many applications the actual analysis and evaluation of texts is done with the help of a lexical resource, that contains information about which individual word can be assigned to what emotion [44]. Assigning all the text elements to the corresponding emotions results in a changing sentiment within the whole text, a document or even a collection of documents.

For the sentiment analysis, a wide range of methodologies has been developed. The analysis of texts often starts with the sentiment analysis on the document level, in order to classify the text as positive or negative [45, 46]. Other approaches produce attribute-value representations, such as frequency of occurrence of different words, called tokens, to allow comparability and simple classification [47]. If sentiment analysis should be more fine grained, methods prefer to segment the document, so that paragraph-based analyzes are possible. A very common approach from sentiment analysis is the lexical evaluation, also known as a bag of words, or dictionary-based approach [48, 49]. This allows a more accurate assessment of the individual components of the text, which were identified in the document pre-processing. The lexical evaluation is used in a variety of sentiment analysis applications [50]. However, some analysis methods use even finer levels of the text to finally assign the polarized text content directly to the entities. In this case, the entity is identified at the sentence-level on the basis of sentence structure (Part-of-Speech). Regarding the sentence structure adjectives, adverbs, verbs and nouns are the typical carrier of sentiment [51, 52].

Some analysis methods also support n-grams, which are common words phrases. In this case, the context is still visible and has not been lost in the tokenizing process. For example, the use of n-grams is examined in the classification of movie reviews from [45] as well as the use of bi- and unigrams which achieved the best results. However, [53] prefers the use of trigrams in the classification of product reviews. N-grams are typically used if there are phrases in textual content which are constantly used as a fixed expression and repeated all over the text. The identification of such patterns within a text is an approach that is also observed by using models of artificial intelligence. In general, a distinction between subjective and objective emotions is made in all approaches and in the study of sentiment [54].

All sentiment analysis approaches have one thing in common - they need a complex and comprehensive document pre-processing. In order to process unstructured data, it requires a specific formatting of the textual data. Therefore, the texts are processed using operations from Natural Language Processing (NLP). For the analysis, this actually means an identification of text levels, usually the endings of sentences, splitting text into n-grams or individual words [53]. Some methods also use text windows, especially those with the help of word associations [55]. Subsequently, the entities are determined on the basis of sentence structure, and finally the text fragments are tokenized. After the application of other NLP techniques, such as removing stop words (typically a list of non meaningful words), the textual data is structured and in a machine-readable form and analytical methods can be applied. A popular approach is the classification of texts in a positive, negative or neutral category, as mentioned before. In this case, the classification methods can for example evaluate the tokenized parts of the original text with a lexical evaluation, summarize it and distinguish on a case basis. Other classification methods prefer to search for textual similarity and distinguish with a distance measure for the identification of category-based cluster centers [56]. Moreover, for the probabilisitc classification the application of bayes classifiers is in a widespread use [57].

Unfortunately, the sentiment analysis still has its limitations. In addition to the limited ability to learn, the analysis faces some unresolved issues in the standard methods. The negation within the sentiment analysis and problems with Word Sense Disambiguation are not completely solved problems yet [50]. By the decomposition and reduction of the text to its essential components (tokens) all relations are lost and are in consequence very difficult to interpret. In particular, irony and satire articles, although recognized as such by the human mind, are very difficult for the sentiment analysis and this issue cannot be solved without a proper artificial intelligence. Often the punctuation gives the reader a hint on a sarcastic remark. However, in most cases the punctuation is removed in the preprocessing of textual content, which can hardly be reconstructed, neither interpreted without them. In addition, in social media abbreviated phrases or terms are often used to communicate. These phrases are unknown to the sentiment tools and must be updated on a regular basis.

Moreover, the emotion in social networks is transported by so-called emoticons. These strings are basically composed of special characters (ASCII) and are usually removed in the preprocessing. On some social media platforms, the use of hashtags is favored. Without adjustments they will be lost in the pre-processing in the standard method, because the number sign (#) is not part of the textual content. In an adapted method for the analysis of social network content they should be classified as an entity, because the real meaning of hashtags is to speak briefly and concisely to a linked topic, person or the current trend. [51] referred to this short text form (e.g. Twitter post with max 140 characters) as a very interesting basis, because it is easy to analyze and authors come directly to the point due to the limitation in characters. Another problem that arose in particular by the increasing popularity of social media, is the so-called opinion spam that is triggered by a variety of comments with emotions and distorts or covers the actual opinion [58]. Intentionally, for a correct analysis false assertions or advertising must be identified as such and removed from the analysis.

## 3. Social media analysis using NoSQL

The use of NoSQL database provides a novel application of social media analysis consisting scalable machine learning and sentiment analysis algorithms for any volume and variety of the data sources and streams. Since SoMABiT provides a mass customized platform that supports importing a wide variety of data sources as well as creating new data jobs for collecting data from social media through streaming APIs, this is a novel application of the social media analysis that supports static and dynamic data integration and analysis. This chapter studies the basic idea as well as already integrated and/or in-progress data sources.

### *3.1. Introducing the SoMABiT concept*

This paper focuses mainly on the scientific background and implementation results of the integrated platform called SoMABiT (Social Media Analysis using Big Data Technology) for the semantic collection of Product Use Information (PUI) through social media. The conceptual idea, is loosely based on the approach proposed by Lovett [37]. In his concept, different data sources of social media are brought together with the support of Hadoop, to be exported into an enterprise environment and visualized in different ways. While the idea proposed in this paper gravitates in a similar direction, the technologies are different, as the core element is a Hadoop based knowledge base (KB). The mashup of different sources creates a tool to support decision making e.g. on a management level or for customers who are about to buy a new product and need support in the decision process. Apache Mahout is to be applied, which has been extended for decision making processes of input queries. For example, "How do customers perceive our newly launched refrigerator on the market?!" or "What is the opinion of <productname>, <productcategory>?!", where <productname> is "Eisbox 1000" and <productcategory> is refrigerator or fridge.

Figure 1 visualizes the basic view of the SoMABiT concept. There are different input data sources of social media like Twitter, blogs, wikis, or product information, e.g. from Amazon, which are addressed via their respective APIs or imported into the NoSQL database. While the product information and evaluation, e.g. from Amazon may not be strictly social media content, the additional source of PUI and rated products allows for data enrichment about the product information in the later mashup procedure. The processing of the acquired data should take place in a cloud environment with a Hadoop based KB. In the case of an inquiry, the request itself is stored with semantic enriched metadata in the database and the process for finding an answer is executed. If there is an answer in the KB, the reply is given to the user and stored together with the request itself. This reprocessing is done with the support of semantic structures, linking the different topics. If there is no solution or answer in the KB available, the cloud logic has to automatically start searching the associated databases for the new information. The reply is stored in the KB again and given as an answer to the user. The inquired and reprocessed facts from the KB will be linked to the relevant statistical numbers to undermine the displayed results. Regarding the storage of social media data, only those APIs or services are considered, which allow the storage of public data provided by an API in regards of copyrights.

Managers, decision makers for the product portfolio, innovation and development are the target users of this application. The mashup interface consists of the numbers, texts, graphics and diagrams, highlighting the most important results discovered from social media such as results of sentiment analysis, KPIs, sales position, tweets where people are currently talking about the product or even comparisons between different countries or brands for analyzing the own products and those of the competitors.

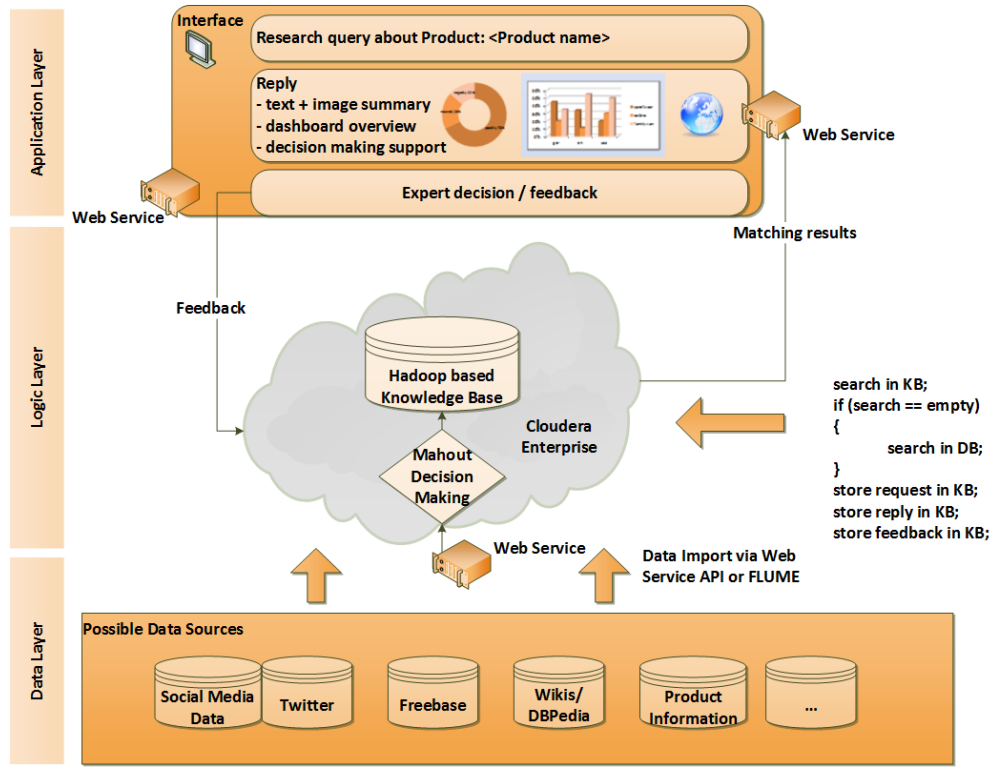

**Figure 1.** The high-level overview of the SoMABiT.

### 3.2. Possible data sources

Possible data may be sourced and streamed from different social media channels and forums, which have officially confirmed the use of the data for scientific and research aims with respect to the data privacy and copyright. Those different data sources that have been studied to be used in the analysis phase, but not yet entirely integrated in the current development are described in the following.

The static data sources such as historical databases and knowledge repositories are being stored in an HDFS structure as data nodes of the virtual infrastructure. Different APIs such as a Twitter Streaming API are being used for the streaming of the data from social media. It has been decided to apply Apache Flume [59] for the collection of data from Twitter streaming API and forwarding to the HDFS deployed KB. The Twitter streaming API outputs are in JSON format, which is not suitable for traditional querying languages. In addition, Amazon Product Advertising API [60] (APA API) is integrated into the KB in order to provide dynamic customer reviews or PUI in general from Amazon. This API will use only the most relevant parts of the product profile on Amazon. The Wikipedia data sources, which are downloaded from sources of open-licensed data published by the Wikimedia Foundation[7] and stored (and replicated) in data nodes of the

Bohlouli et al     9

system, cover the statistics specially website traffics on the specific topic which can proof the popularity of the topic in developed dashboard. The Wikipedia data sources consist of 2.5 million wiki pages with about 2.5 TB in size. The data is stored in XML forms and being processed via Hive queries. Hive [13] is an Apache project for data warehousing, ad-hoc querying and data summarization in a Hadoop ecosystem.

Freebase data dump[8], which is one of the world's largest online and open information source databases, may be integrated in the project dashboard, as it consists of the information of over 37 million entities. The complete data is stored in the data nodes in RDF form and is being updated using the Freebase Topic API. In addition to the proposed databases which have been retrieved from freely available and officially permitted sources, few more public data sets on Amazon Web Services (AWS) are under study to be integrated into the system. One of the complex steps in the data streaming functions is to automatically move the data to HDFS without manual intervention. To this aim, Apache Flume facilitates the data pipeline.

### 3.3. Envisioned user interface

For visualizing the results of the aforementioned analysis process, the singular results of different social media sources are brought together in a mashup to provide an overview supporting decision making for the further product lifecycle of the product. In the given example (section 3.1) of a refrigerator or fridge (exemplarily called "Eisbox 1000"), the manager can see the web sentiment of the product, current tweets, KPIs from Twitter, entries from the company blog, product information or evaluation from Amazon as well as related terms, wiki entries and countries where people talk mostly about the product. With the query just asking for the <productname> and <productcategory>, it is furthermore possible to get a short survey of a competitor's product. The single diagrams are schematically displayed in figure 2.

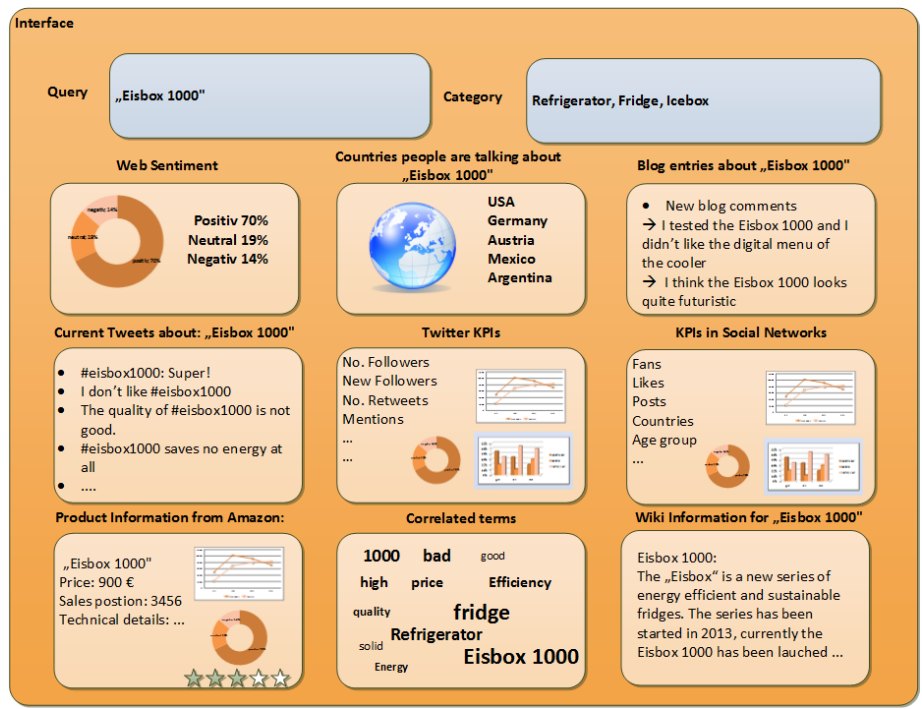

**Figure 2.** Product information collection – Scheme of result interface.

## 4. SoMABiT realisation

The realization of the SoMABiT proofs the novelty of the application to find out whether integrated machine learning and sentiment analysis algorithms are efficient enough to be compared with traditional similar applications. In this way, the test and evaluation phase ensures the velocity of the system as well. The following chapter describes the functional architecture, practical details of sentiment analysis in MapReduce by means of pseudocodes and also provides the analysis of the results and findings.

### 4.1. Functional architecture

The SoMABiT architecture consists of three components: (1) SoMABiT-Server, (2) SoMABiT-WebServlet and (3) GUI for interacting with end users. The SoMABiT-Server is used in a Cloudera cluster system and has the important task to provide the data required for analysis. The component can connect to the social media like Twitter and thereby initiate a data retrieval via an API. The information obtained in a form of tweets is stored in the HBase database. In this process, the SoMABiT-Server itself configures the required tables. Furthermore, it converts the extracted data into a unified format for storage. In addition, other data sources can be imported into the system in a form of CSV files via universal interfaces. This is intended to ensure that other social media (or data sources) can be integrated into the system for which no separate streaming interface is implemented or provided by the social media.

To determine the selection of data, a system of so-called data jobs is used. These data jobs are stored in a separate database (MySQL). At regular intervals, the SoMABiT-Server requests information from the database if new data jobs have been created by the user, which sources are queried, and at which time the data should be imported. The communication between SoMABiT-Server and the database takes place via the JDBC interface. The SoMABiT-WebServlet is ideally installed and set up on the separate server with the MySQL database that stores the information about data jobs. This service includes functions for creating, modifying, and deleting data jobs and users. It also represents the main function to provide analytical support to the dataset of an HBase database. For this function -equivalent to the data jobs- the opportunity to create so-called MapReduce jobs is given. They process the data from HBase tables according to their analysis purpose.

The SoMABiT-WebServlet handles the complete processing of analytical data and therefore offers the basic functionality of the service. The service communicates, as mentioned before, with the SoMABiT-Server through the JDBC API with a MySQL database, and with the HBase API to an HBase database. For interaction with a user, a web browser-based application is used as a graphical user interface. End users of the system can access and control SoMABiT-Server and SoMABiT-WebServlet via web-browser. This is mainly done by creating data and MapReduce jobs.

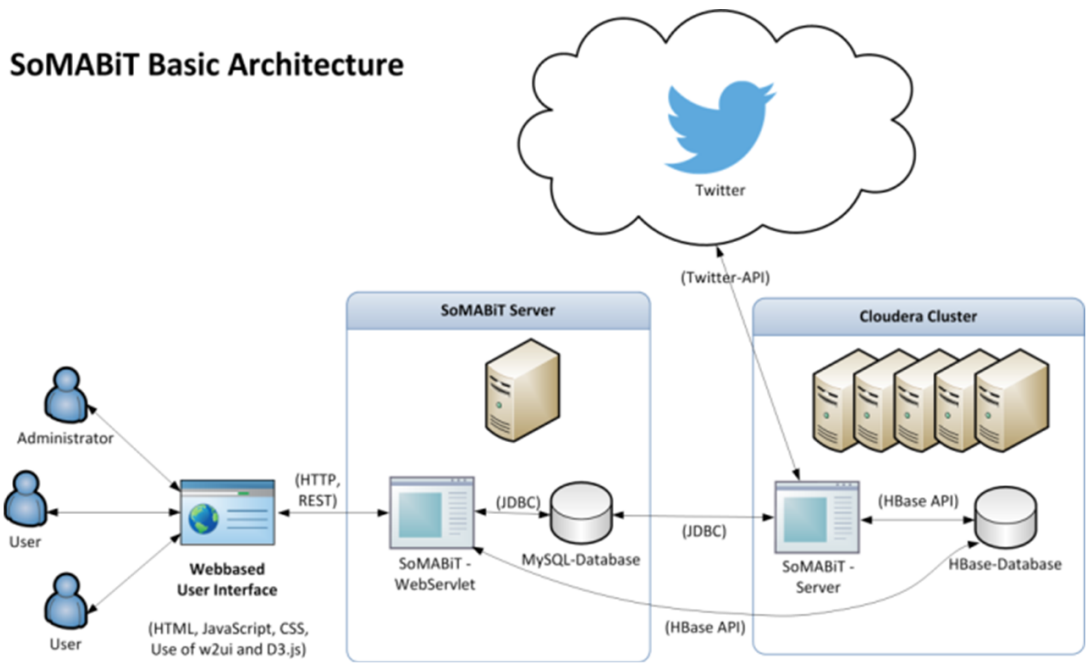

**Figure 3.** The functional architecture of the SoMABiT system.

A distinction is made between two groups of users: "User" and "Administrators". While the normal users have the ability to generate reports on the data stocks, the administrators have more features: They are allowed to create and run data jobs, to store datasets into the system. In addition, an administrator can administer the HBase database and remove individual data sets from the system. Communication with the SoMABiT WebServlet runs over the HTTP protocol, the returns take place in the form of JSON-formatted data. Requests have already been converted into the JSON format and sent to the WebServlet. This can be determined based on parameters, which define the functions to execute and what data needs to be returned. The WebServlet can distinguish whether the MySQL database records are read, altered or created and performs the desired functions. The actual logic is thus anchored in the WebServlet while the user interface is used just for communication and visualization of content.

Figure 3 illustrates the functional architecture of the system. Theoretically, it is possible to operate the application components on only one system. However, the decision about this is made at a later operation of the entire application. Depending on the availability, individual computer resources, the SoMABiT application can be scaled by dividing the components SoMABiT-Server and SoMABiT-WebServlet.

### 4.2. Technical architecture

The virtual infrastructure used for the implementation and development of the project consists of 20 quad-core virtual machines (VMs) with 500 GB storage and 4 GB RAM memory for each. As a core manager of the used virtual infrastructure, the Cloudera distribution, including Apache Hadoop (CDH4) is configured. The technical specifications of the virtual infrastructure are illustrated in Figure 4. The front node of the system is named "NameNode1" and uses the compute nodes (CNode10-29) for the processing of the complex and distributed data. Databases are replicated at the rate of 3 and support fault tolerant data storage. A Zookeeper service is replicated over multiple machines and provides the centralized infrastructure which stores the information about the status of the entire cluster in the form of log files and facilitates the synchronization across the cluster. The data is replicated between compute nodes. Hue and Hive [13] provide proper tools for the client access. Hue is a file browser service which supports navigation and files/folders basic functions in HDFS such as creation, deletion, search and smart navigation. Hive provides the querying language in the Hadoop ecosystem. User requests and queries are being processed through Hive.

The SoMABiT-Server runs as a normal Java application. In contrast the SoMABiT-WebServlet requires a corresponding servlet container. The browser-based user interface is created using HTML and JavaScript. The advantage of dividing the application into three different components is that a high range of scalability can be achieved. The SoMABiT-Server runs on the Cloudera cluster system and is responsible for the provision of data. The SoMABiT-Webservlet and the user interface can be outsourced to other servers, so that these applications do not need to burden them with other objects of the cluster. The user interface takes advantage of the additional resources of the client PC for the visualization. The SoMABiT-Server must be able to recognize when new data jobs have been created and how they should be performed.

The SoMABiT prototype's first step is to connect to a MySQL database which holds the data jobs in a table. The SoMABiT-Server retrieves this data and checks whether a new data job has been created. If this is the case, it is checked whether the job data is to be executed immediately or if it is scheduled at a later time. For the execution it must be determined at first from which data source the data has to be imported. Available are the Twitter network and the import from CSV files.

Appropriately, a corresponding target table in the HBase database is created. Next, the reading of the file is specified. This is done sequentially. After each reading, the conversion of the data in a suitable format and storage takes place. If the file has been read as a whole, it is closed and a protocol about the data import is created. If data from the Twitter network are read, first a target key-value pair is created in the HBase database. Then, using the OAuth process to connect to Twitter, the retrieval of individual tweets is being started. These will also be placed in a suitable format and stored in the HBase table. After each received tweet, it is checked whether anymore should be retrieved. This is defined by the data job that must provide information in the form of retrieval time or number of records. Once the import has been completed, the connection is closed and an import log is created.

The processing of MapReduce jobs takes place analogously to the processing of data jobs by the SoMABiT-Server. Once the server has connected to the MySQL server, it checks whether new MapReduce jobs have been created. If this is the case, they are executed immediately, contrary to data jobs. In the following the MapReduce job determines which algorithm has to be used for data transformation. With this information, the SoMABiT-Server can log on to an HBase database and create an appropriately formatted target table. Subsequently, the application of the desired MapReduce algorithm on the selected source database or table takes place. Here, the MapReduce job writes the recalculated records in the target table. Finally, a log is created. To describe the operation of the data jobs the different states are described in section 4.3. After creation of a data job the status "JOB_GENERATED" is set.

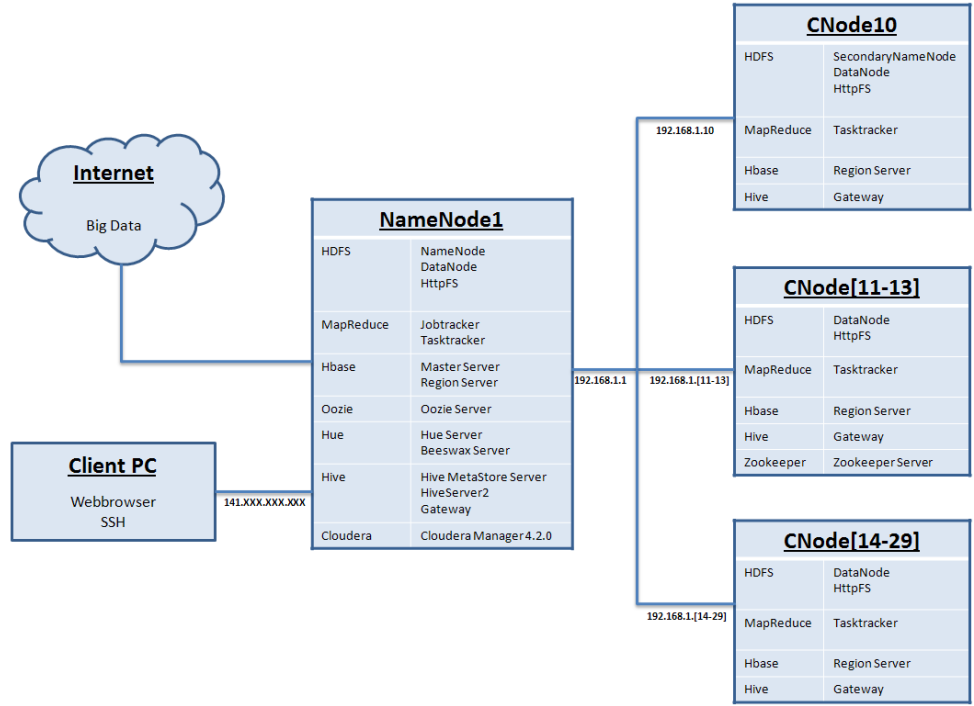

**Figure 4.** The technical configuration of the SoMABiT virtual infrastructure using Cloudera Enterprise.

In consequence the SoMABiT-Server checks at specified intervals the MySQL database for new data jobs. If it detects a new one, it sets the status to "WAITING". Following this it checks whether the data job has to be executed immediately or at a specified time and changes the status to "RUNNING". The data import begins. After its completion the status is either set to "SUCCESS" or in a negative case to "ERROR". Analogous to the data jobs a job system for the MapReduce jobs can be used. However, these are to be executed directly, so no check on timelines must be made and a newly generated MapReduce job changes directly to the status "RUNNING".

The SoMABiT-WebServlet is used to communicate with a user in the form of a user interface for data input and output. When inquiries are made to the system through the user interface, it is first checked whether the user can be authenticated. The user may either request storage or retrieval of data. If a record is saved, this is done by the SoMABiT-WebServlet and a corresponding return is generated that is sent back to the user interface. When querying data the required information is presented by returning it to the requesting client.

### 4.3. Sentiment analysis algorithm based on MapReduce

In the analysis method for text classification by SoMABiT a lexical resource is used with the aid of two separate word lists of pre-classified positive and negative words with a numerical value on the interval scale [-1,1]. In the following part the algorithm for the SoMABiT sentiment classification is introduced with pseudocode listings in three steps. In the first step keywords are iteratively searched in the texts, which are stored in the table entries in the HBase database. If a keyword was found, the record is stored in a new table, otherwise it is skipped (Figure 5).

```
// 1. Set variables
set hbase_table = $table
set filter_column = $column
set filter_keyword = $keyword

// 2. Mapping phase
while $hbase_table != EOF
if $actual_record.filter_column.contains(keyword)
put $actual_record in $newtable
endif
endwhile

// 3. Reducing phase
// Nothing to do here, the records are already stored in the mapping phase.
```

**Figure 5.** The map and reduce algorithm of the filtering.

In the second step the preprocessing of the textual content is realized by the implementation in the mapping phase. After tokenizing and the removal of stopwords a frequency table with all tokens from the table entry is created by the word count algorithm (Figure 6).

```
// 1. Set variables
set hbase_table = $table

// 2. Mapping phase
while $hbase_table != EOF
array_words = tokenize(hbase_table.column)
for each $word in array_words
put ($word, 1) in $newtable
endwhile
endwhile

// 3. Reducing phase
while $newtable != EOF
if $newtable.word != $lastword
$newtable.count++
endif
endwhile
```

**Figure 6.** The map and reduce algorithm of the wordcount.

In the third step, the lexical resource is used to classify the texts. The entries of the new table from second step are iteratively compared with the positive and negative entries of the word lists. If the words match, the numeric value is stored. A numeric average value is calculated at the end (Figure 7). In conclusion, if the average value is above 0 the keyword is associated with a positive sentiment. If the average value is below 0 the sentiment belonging to the keyword is negative.

```
// 1. Set variables
set positive_words_array = loadfile(positive_words)
set negative_words_array = loadfile(negative_words)

// 2. Mapping phase
while $hbase_table != EOF
array_words = tokenize(hbase_table.column)
for each $word in array_words
if $word in positive_words_array
put ($word, positive_words_array_value) in $newtable
elseif $word in negative_words_array
put ($word, negative_words_array_value) in $newtable
endif

endwhile
endwhile

// 3. Reducing phase
while $newtable != EOF
$mean_value = 0.00
$counter = 0
if $newtable.word != $lastword
counter++
$mean_value = mean_value + $newtable.word_value
endif

$mean_value = $mean_value/$counter

put $newtable.value = $mean_value
endwhile
```

**Figure 7.** The map and reduce algorithm of the sentiment analysis.

### 4.4. User interface and evaluation of the results

The SoMABiT dashboard currently offers five different visualization types in the form of (1) charttable, (2) bubble chart, (3) circle packing, (4) bar chart and (5) line. The charttable creates an HTML-formatted output for displaying rows and columns of the data in an existing HBase table. This visualization option is used to display specific text information from the selected source(s) table. The bubble chart option generates individual circles from given data, each representing a specific element in the form of a selected source column. Circle packing is the most powerful visualization type. It is particularly suitable for data, which have been generated with a classification MapReduce job. It also provides a chart with several circles, additionally representing a group or a class. In addition, by the color of each item it is expressed whether the sentiment of a text is evaluated positive or negative. More green elements are positive, while more red elements are considered negative. Figure 8 displays this kind of visualization based on test data categorizing between different countries an the mentioning of different colors.

In order to test the functionality of the system and proof achieved results, SoMABiT has been tested in three different use case scenarios. The first scenario targets the automotive industry, where the producer is able to discover knowledge from social media about market needs and customer feedback (e.g. comments) for specific cars or their features. Furthermore, it is helpful to identify the problems with individual vehicles or an entire brand so that a manufacturer can counteract by corresponding improvements in the next generation of the product. This also helps industries in order to identify the most preferred products and/or their features in the market and orient themselves in that direction. This scenario also discovers the regional preferences, for example the most preferred colors of the cars in the specific country.

For this, there must be a categorization of countries. Alternatively, this also provides an insight about specific products or services to the end users (customers).

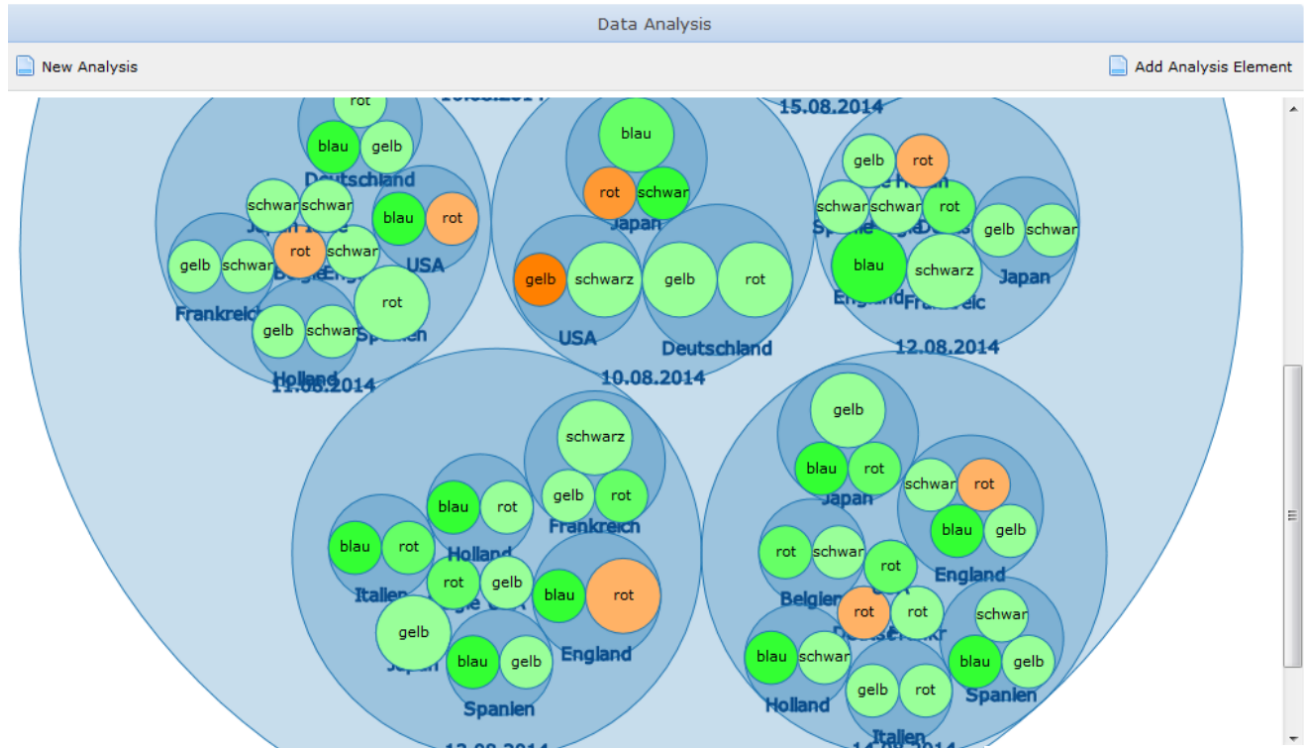

**Figure 8.** Product information collection – Scheme of result interface.

The second use scenario targets the football area in which the authors have tested the system in the time-frame of the FIFA World Cup 2014 in Brazil. In this scenario, social media supports discovering the information specially the sentiments about any specific football player, team and/or country. The feelings of the people expressed in a written form as well as discussions about specific matches can help the system to provide useful information in this regard. The application provides, for example, the possibility to determine on individual matches which players either have been perceived more positively or negatively. This in turn can be determined and graphically visualized by using the classification algorithm. Thus, it can be determined whether a particular perception is reflected in all countries or whether regional differences exist. As stated earlier, the SoMABiT system provides different types of the visualizations in a form of the charts. Figures 9.a and 9.b show the bar and line chart defined in this scenario. There is also a possibility to use other chart types integrated in the system for any scenario. In the bar chart, users can compare different categories like countries or football players based on the sentiment values such as which team has a potential to be a winner. The line chart provides the sentiment over the time or the development of the sentiment like which team had a great improvement during the specific time period. This can for example visualize the time period that a specific trainer has been employed by the team which identifies whether the trainer was considered successful or not based on the opinion of the football fans.

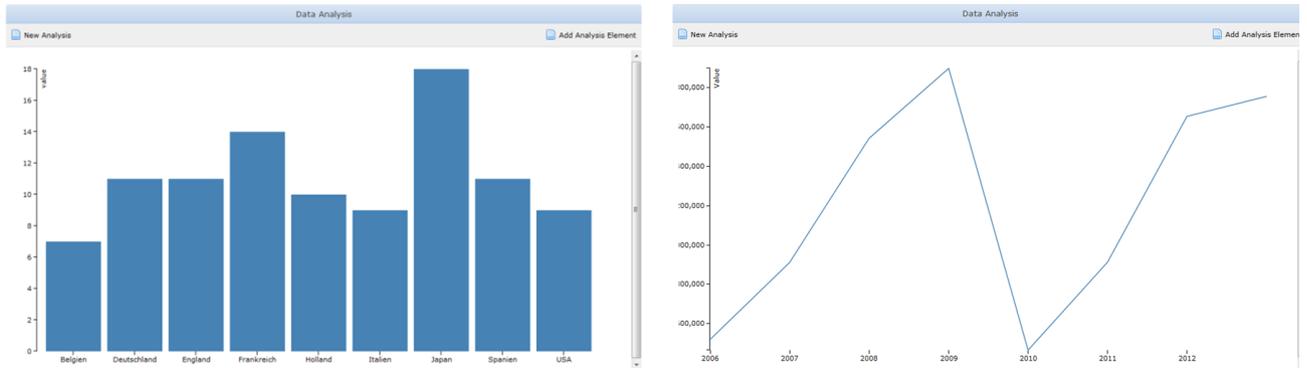

**Figure 9.** Graphical Visualization of the football scenario through the information collected from social media. The bar chart indicates comparing of the national football teams based on the sentiment value and the line chart illustrates the development of the sentimental value for specific football player over the time.

The third example is applied in the area of politics. Since this is a very wide field, the test scenario was limited to few hot political topics and discussions at the time of testing the system. These currently include the prevailing conflict in Ukraine and the military actions in the Gaza Strip between Israel and Palestine. In this scenario, political news and discussions provide the information for the sentiment analysis of different political issues such as what is being decided in the voting, or which political party has more supporters in which regions and for which justifications. For instance, figure 10 provides the visualized insight about the posted regional news about the Ukraine conflicts in specific countries.

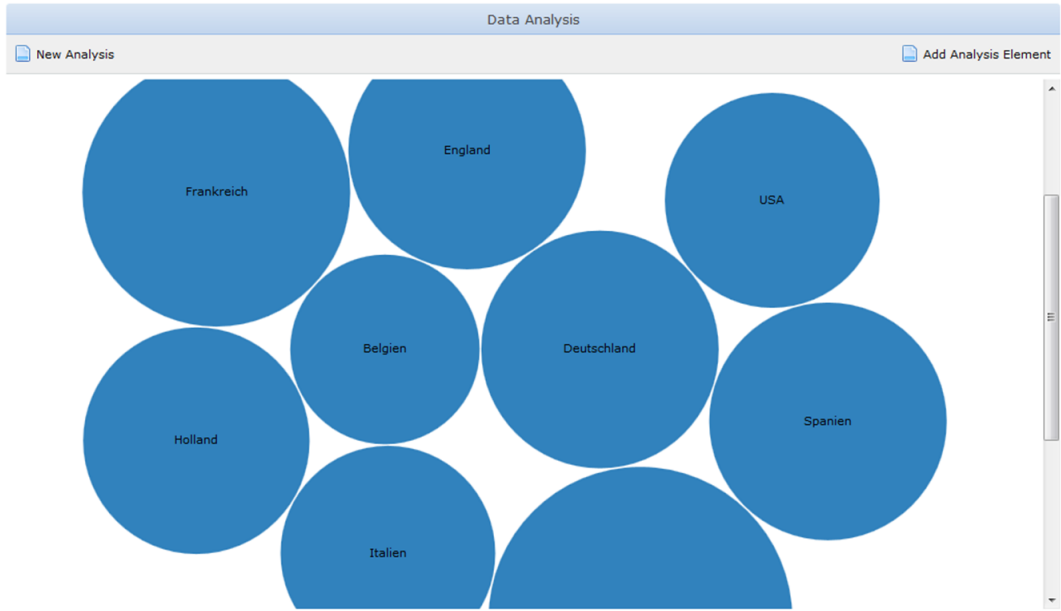

**Figure 10.** Graphical visualization about the total number of posted regional news about the Ukraine conflict.

In order to evaluate the results, the project test phase consists of four steps as functional, non-functional, usability and performance testing. Functional testing is a use case based testing and covers the deviations from basic flow and wrong user inputs. The ease of using the system is being assessed in the non-functional testing. Unit tests and non-functional requirements are parts of this phase. In the usability test, a sort of indicators have been defined which ensures the quality of SoMABiT platform. The inputs and feedback for the corrective measures have been collected. Additionally, the scientific results and semantics of the project have been tested based on the real world scenarios in earlier stated use cases.

## 5. Conclusion and future work

The conceptual idea, scientific focus, realization results and the collection of the product use information via different, multi-structured and disparate social media sources and streams have been discussed in this paper. The core and key innovation of the platform is the use of cloud deployed NoSQL database technology for development of the project knowledge base and visualization of the knowledge discovered from social media in a form that supports end-users and domain experts in making right decisions and/or creating market oriented novel product ideas. In addition, the SoMABiT, in comparison to the state-of-the-art technologies and research outcomes, is a novel application of the big data to provide a scalable decision support platform (sentiment analysis and classification algorithms) for any volume and variety of disparate data sources and streams. This facilitates the analysis of the dynamic and static data. The dynamic data collection is being handled through streams APIs like Twitter streaming API and the static data collection is supported though importing the data sources in a form of CSV files into the system. The SoMABiT analysis phase visualizes the achieved results of both data collection types (e.g. dynamic and static) in a unique output to the end users. The Cloudera Manager builds multi-node CDH cluster with configured Hadoop and HBase integration. The project database is based on the HBase technology.

As a future work, it is planned to integrate more individual databases in order to provide more complete product specific information packages in the mashup tool. Consideration of the scaling up and adding of few more hardware resources could increase the response time of the software tool. In this regard, the response time analysis could efficiently increase the performance of the system. The consideration of a learning component, which can train the functionality of the mashup tool by analyzing the user feedback about the total performance of a system and may significantly increase the accuracy of the system.

The authors studied the potentials for integration of semantic technologies in the frame of the SoMABiT project. As stated earlier, semantic technologies are being integrated in the logical layer of the SoMABiT in order to connect the different data sources and topics with each other for improved knowledge visualization. Meta data provided by the sources, e.g. the hashtag (#) information of Twitter plays a key role in this categorization and connection of data. Based on the idea of the LOD cloud[6,9], there will be an analysis, how the data layer may be enriched, with already semantically structured RDF data, that may be connected to the application via direct data import (upload) or a provided API via SPARQL queries.

An efficient management of the human resources as well as allocation of the right employees with the right expertise to the right job roles improves significantly the performance of the organizations. To this aim, identification of the competence gaps is the first and most important step. In order to fulfill the identified gaps, organizations should assess the employees' competences as well and find the best fitting candidate to the this competence gap. According to the reference competence pyramid, that authors developed in the frame of the European research project [61, 62], called ComProFITS[10], social competences are one of the most important competences in the assessment. In the current research project, the 360 degree assessment method is being used to evaluate the social competences of the employees. SoMABiT platform has a great novelty and productivity to be used in this project in order to automatically and actively discover the social competences of employees through their involvement in the social media.

Additionally, SoMaBiT is going to be extended as a cloud-based Knowledge Integration (KI) platform [63] which collects information from social media, especially Twitter, and provides further analytics as a service. Accordingly, an extension of the SoMABiT as a cloud service in the frame of XaaS (SoMaaS: Social Media analysis as a Service) will improve the elasticity of the system for further processing and resource demands.

**Notes**

1. About Twitter Inc., https://about.twitter.com/company (accessed June 2015).
2. Lea, Wendy, 'Infographic: The Potential of Big Data', http://blog.getsatisfaction.com/2011/07/13/big-data/?view=socialstudies (2013, accessed June 2015).
3. Youtube Statistics, https://www.youtube.com/yt/press/statistics.html (accessed June 2015).
4. Semantic Web Standards, http://semanticweb.org/wiki/Semantic_Web_standards (accessed June 2015).
5. The World Wide Web Consortium (W3C), http://www.w3.org/ (accessed June 2015).
6. Linked Data – Connect Distributed Data across the Web, http://linkeddata.org/ (accessed June 2015).
7. Research Data: Meta, Wikimedia Foundation , https://meta.wikimedia.org/wiki/Research:Data (accessed June 2015).
8. Data Dumps - Freebase API, https://developers.google.com/freebase/data (accessed June 2015)
9. Linked Open Data Could, http://lod-cloud.net/versions/2014-08-30/lod-cloud.svg, (accessed June 2015)

10. European Research Project: Competence Profiling framework for IT sector in Spain (ComProFITS), www.comprofits.eu (2013, accessed June 2015).

**Funding**

This research received no specific grant from any funding agency in the public, commercial or not-for-profit sectors.